\documentclass[aps, superscriptaddress, twocolumn, floatfix,showpacs]{revtex4}

\usepackage{graphicx}
\usepackage{amsmath, amssymb}
\usepackage[usenames]{color}
\begin{document}

\title{Droplet and cluster formation in freely falling granular streams}

\author{Scott R. Waitukaitis}
\affiliation{ Department of Physics, The University of Chicago, Chicago, IL 60637}
\affiliation{James Franck Institute, The University of Chicago, Chicago, IL 60637}

\author{Helge F. Gr\"utjen}
\altaffiliation{current address:  Centre for Mathematical Sciences, University of Cambridge, Cambridge CB3 0WA}
\affiliation{James Franck Institute, The University of Chicago, Chicago, IL 60637}

\author{John R. Royer}
\altaffiliation{ current address:  Center for Soft Matter Research, Department of Physics, New York University, New York, New York 10003}
\affiliation{ Department of Physics, The University of Chicago, Chicago, IL 60637}
\affiliation{James Franck Institute, The University of Chicago, Chicago, IL 60637}

\author{Heinrich M. Jaeger}
\affiliation{ Department of Physics, The University of Chicago, Chicago, IL 60637}
\affiliation{James Franck Institute, The University of Chicago, Chicago, IL 60637}

\date{ \today}

\begin{abstract}
Particle beams are important tools for probing atomic and molecular interactions. Here we demonstrate that particle beams also offer a unique opportunity to investigate interactions in macroscopic systems, such as granular media.  Motivated by recent experiments on streams of grains that exhibit liquid-like breakup into droplets, we use molecular dynamics simulations to investigate the evolution of a dense stream of macroscopic spheres accelerating out of an opening at the bottom of a reservoir. We show how nanoscale details associated with energy dissipation during  collisions modify the stream's macroscopic behavior. We find that inelastic collisions collimate the stream, while the presence of short-range attractive interactions drives structure formation.  Parameterizing the collision dynamics by the coefficient of restitution (i.e., the ratio of relative  velocities before and after impact) and the strength of the cohesive interaction, we map out a spectrum of behaviors that ranges from gas-like jets in which all grains drift apart to liquid-like streams that break into large droplets containing hundreds of grains.  We also find a new, intermediate regime in which small aggregates form by  capture from the gas phase, similar to what can be observed in molecular beams.  Our results show that nearly all aspects of stream behavior are closely related to the velocity gradient associated with vertical free fall.  Led by this observation, we propose a simple energy balance model to explain the droplet formation process.  The qualitative as well as many quantitative features of the simulations and the model compare well with available experimental data and provide a first quantitative measure of the role of attractions in freely cooling granular streams.
\end{abstract}


\maketitle

Granular material consists of macroscopic, solid particles interacting predominantly via contact forces \cite{Jaeger:1996p1392,Goldhirsch:2003p1343, Duran:2001p2846}.  In modeling granular flows these interactions are typically taken as purely repulsive. However, there are important circumstances, from agglomeration in fluidized particle beds to dust accretion in proto-planetary discs, where attractions can compete with the particle weight or with forces produced by particle collisions and then lead to the formation of stable granular clusters \cite{Castellanos:2005p2595,Spahn:2004p2498,Ulrich:2009p2669,Weber:2006p2638}.  In dry, nominally free-flowing granular material the attractions are small and short-ranged, and are usually associated with van der Waals forces or capillary bridges due to a few layers of adsorbed molecules \cite{Visser:1989p2640,Weber:2006p2638}.  To measure the resulting residual cohesion between grains, atomic force microscopy (AFM) has proven useful \cite{Tykhoniuk:2007p556,Jones:2003p1678,Castellanos:2005p2595}, but this technique mimics the static limit of central, head-on collisions.  Under dynamic conditions, energy is also dissipated due to the inelastic nature of collisions, including rolling and sliding friction during non-central impacts \cite{Brilliantov:2007p2661,Brilliantov:2004p1671,Marshall:2009p2664,Spahn:2004p2498,Zippelius:2006p2657}.  Simulations have started to address the competition between inelasticity and cohesion during collisions in freely cooling granular systems, but so far have focused on the limit of a dilute granular `gas'  \cite{Brilliantov:2006p2662,Marshall:2009p2664,Ulrich:2009p2669}. Little is known about the complex dynamics that lead to clustering in the dense limit, where many weakly cohesive particles collide and interact in rapid succession.

New possibilities to tackle this problem have emerged from recent experiments on freely falling granular streams \cite{Amarouchene:2008p2673,Mobius:2006p566,Royer:2009p41}. In these systems, a dense stream of particles emerges from a small opening at the bottom of a reservoir and is accelerated downward by gravity. During vertical free fall, particle interactions lead to spatial inhomogeneities that can be detected downstream.  This makes it possible to observe the effects of attractive forces as small as nanoNewtons between macroscopic grains \cite{Royer:2009p41}.  Such forces constitute an effective `surface tension' that can affect the bulk dynamics of the stream.  In principle, the sensitivity afforded by such experiments should provide a means to delineate cohesive from purely collisional contributions to structure formation. This is because in the accelerated, co-moving frame the stream undergoes extensional flow and particles initially touching will eventually separate along the axial direction unless they are held together by attractive forces.  This is similar to what happens during free expansion of molecular jets \cite{Bruch:2002p2686,Harms:1997p2674}. The details of cluster evolution, shape, and size therefore contain information about the inter-particle interactions.  However, to access this information, a connection between cluster properties and relevant particle parameters must first be established.  

As a first step, we report here on a set of systematic simulations designed to differentiate between the roles of inelasticity and cohesion in driving droplet formation or clustering.  Such simulations can give access to aspects difficult to probe experimentally, such as the local particle configurations and dynamics in the interior of the stream.  Furthermore, they can explore parameter ranges that so far have not been tested in experiments. 

Covering large regions in the cohesion-inelasticity parameter space, we  not only reproduce qualitatively the observed experimental findings, but gain new insights regarding the roles of inelasticity and cohesion as well as the mechanism through which droplet formation occurs. In terms of the gross behavior, we are able to delineate regimes ranging from sprays of isolated particles to aggregation into clusters by collision and capture to  break-up into droplets as in Fig.~\ref{fig:clusters}(d) or even larger, solid-like chunks.   On a much more local scale, by tracking the average contact number of particles in the stream, we demonstrate that clustering and droplet formation are signatures of attractive forces, while inelasticity plays only a secondary role by helping to collimate the stream.  We also show that the average network size correlates directly with the cohesion strength.  This opens up unique possibilities to quantify cohesion by simply measuring the network size and, more generally, to use granular streams as sensitive probes of inter-particle attractive interactions.

\section{Results and Discussion}

\begin{figure}
\includegraphics[width=.48\textwidth ]{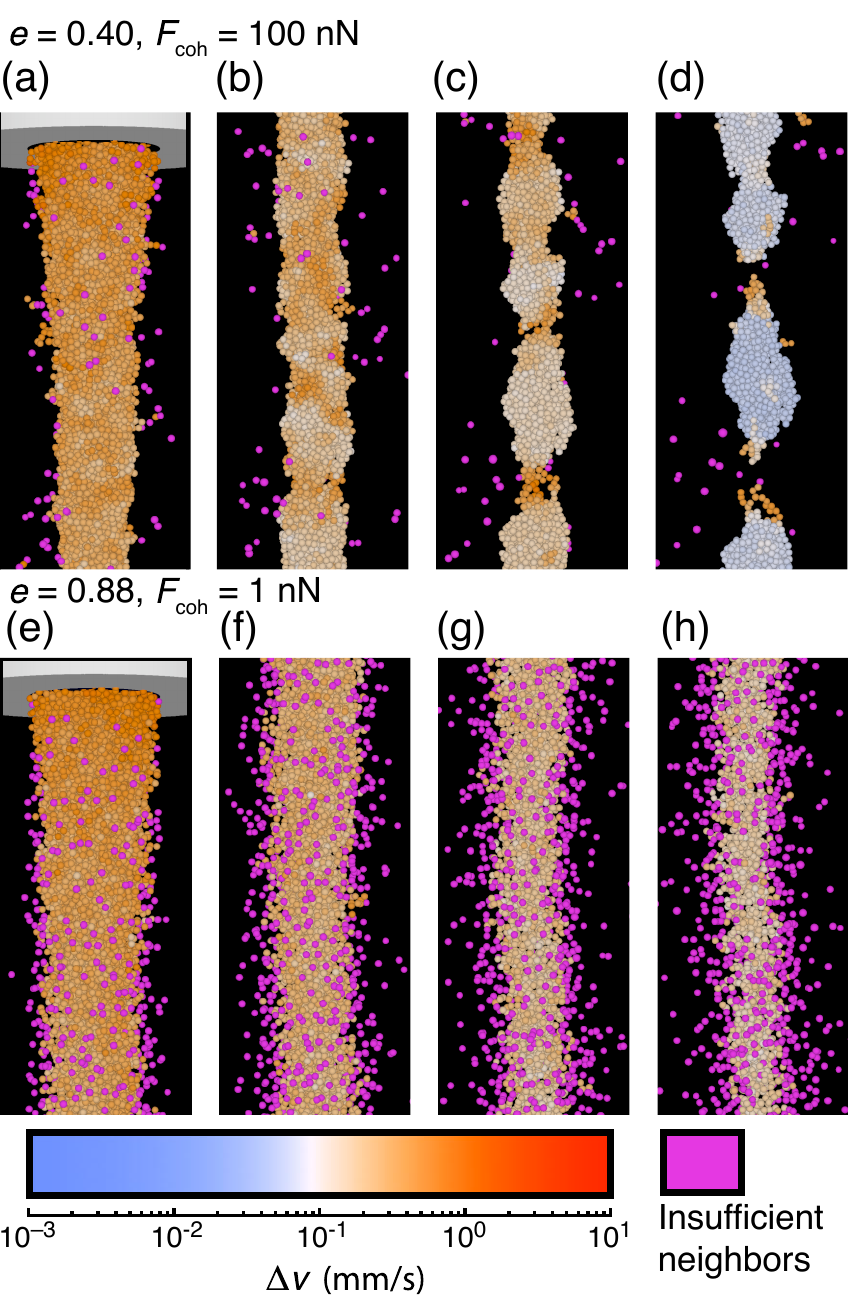}
\caption{(Color online) Snapshots from simulations of granular streams consisting of $d=$~100 $\mu$m spheres freely falling out of a $D_0=$~2.0 mm aperture.   (a)-(d) Coefficient of restitution $e = $~0.40 and cohesive strength $F_{coh} = $~100 nN.   (e)-(h)  $e = $~0.88 and $F_{coh} = $~1 nN. The images follow the grains in the co-moving frame and are taken just below the aperture [(a),(e)], and at distances $z=$~0.7~cm [(b),(f)], 1.9 cm [(c),(g)] and $z=$~3.6 cm [(d),(h)] from the aperture to the top of each frame. Grains are color coded according to their rms velocity difference with neighboring grains within a distance of 1.5$d$. Isolated grains are shown in pink.  Images generated via \cite{raster}.  For movies see \cite{epaps_note2}.}
\label{fig:clusters}
\end{figure}

\begin{figure*}
\includegraphics[width=.95\textwidth]{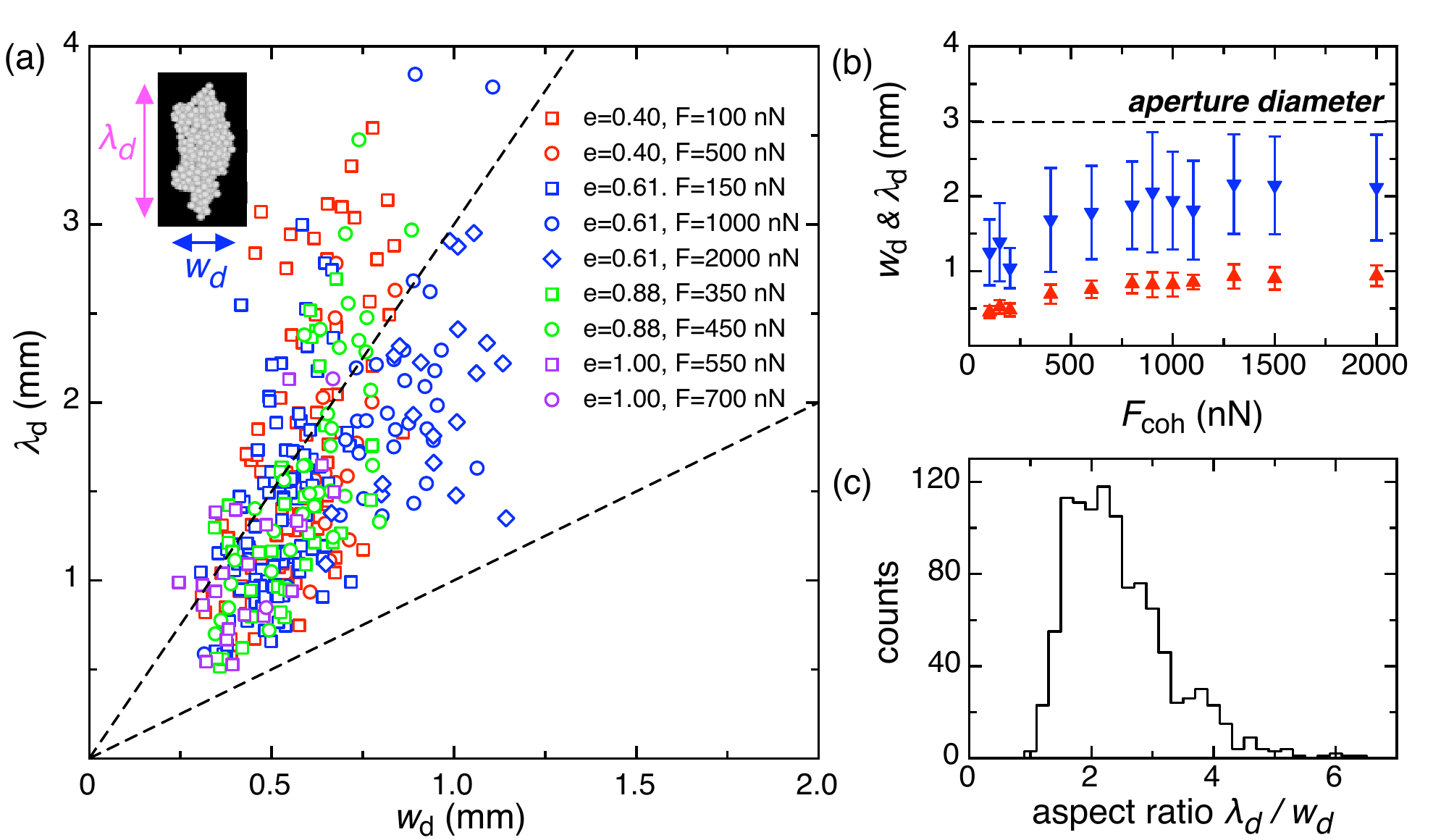}
\caption{(Color online)  Droplet shapes for $d=$~200 $\mu$m spheres freely falling from a $D_0=$~3.0 mm aperture. (a) Droplet height $\lambda_d$ vs. width $w_d$ for simulations in droplet forming regime.  Only a sampling of simulations is shown here to avoid clutter [data for \emph{all} droplet forming simulations are included in Fig.~\ref{fig:aspect}(c)].  Inset demonstrates what is meant by width $w_d$ and height $\lambda_d$. (b)  Droplet width $w_d$ (red) and height $\lambda_d$ (blue) vs. $F_{coh}$ for $e=0.61$.  Error bars show standard deviation of measured $\lambda_d$ and $w_d$. Aperture diameter is indicated by dashed line.  (c)  Distribution of aspect ratios $\lambda_d/w_d$ for all droplet forming streams.} 
\label{fig:aspect}
\end{figure*}

\begin{figure}[t]
\includegraphics[width=.48\textwidth]{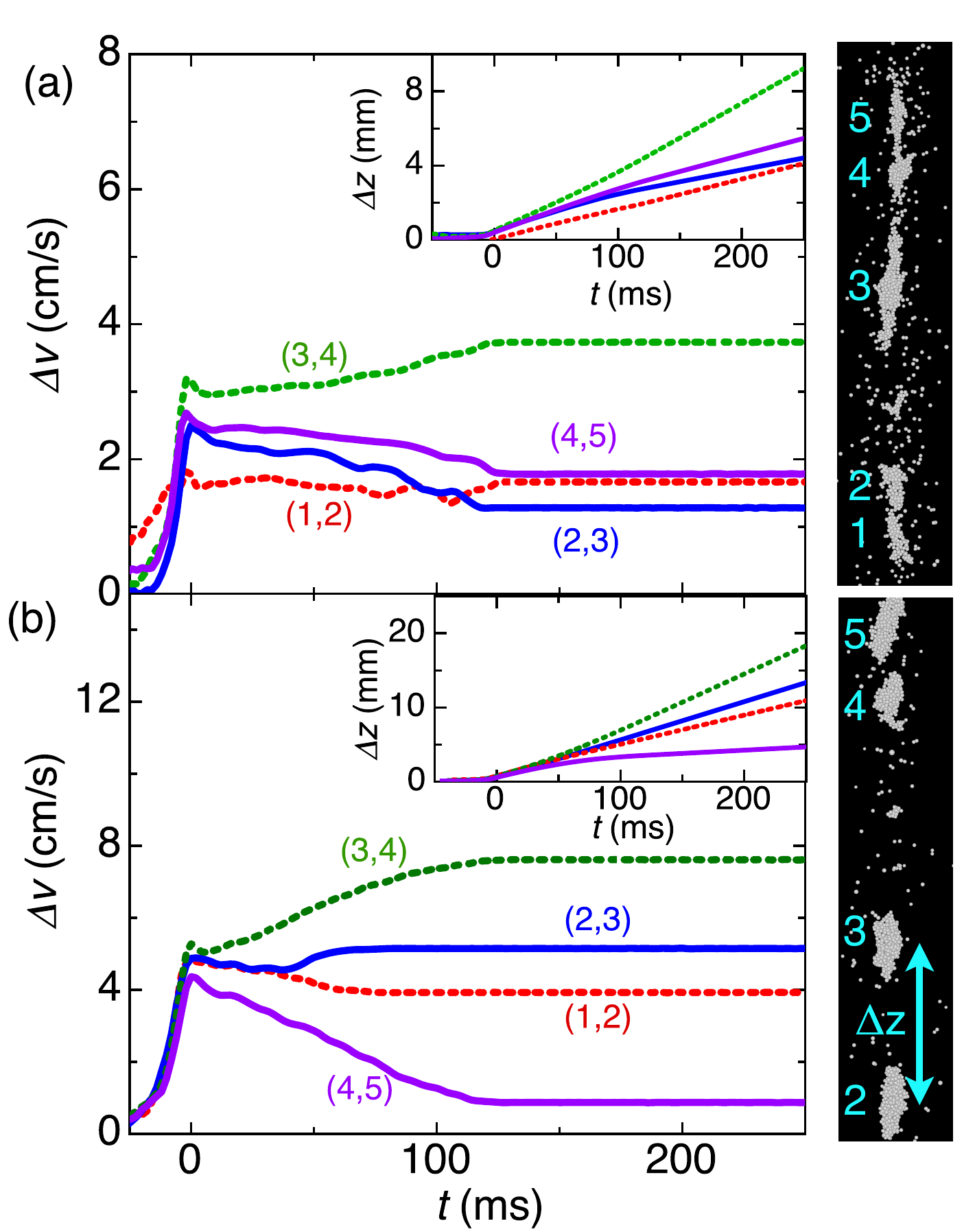}
\caption{(Color online) Droplet separation dynamics.  Data are for the same grain diameter and aperture size as in  Fig.~\ref{fig:aspect}.  Distance between the center of mass $\Delta z(t)$ (insets) and separation velocity $\Delta v(t) \equiv d(\Delta v)/dt$ for four pairs of adjacent droplets from streams with $e = 0.61$ and (a) $F_{coh} =$~400~nN and (b) $F_{coh} = $~1100~nN.   For each pair, the the time $t$ is offset so that at $t=0$ the center of the pair of droplets passes through the aperture ($z=0$).  Images to the right show the pairs of droplets tracked here at a depth $z=$~18.4~cm below the aperture ($t \sim $ 180 ms).} 
\label{fig:dv_dz}
\end{figure}

\begin{figure*}
\includegraphics[width=.95\textwidth]{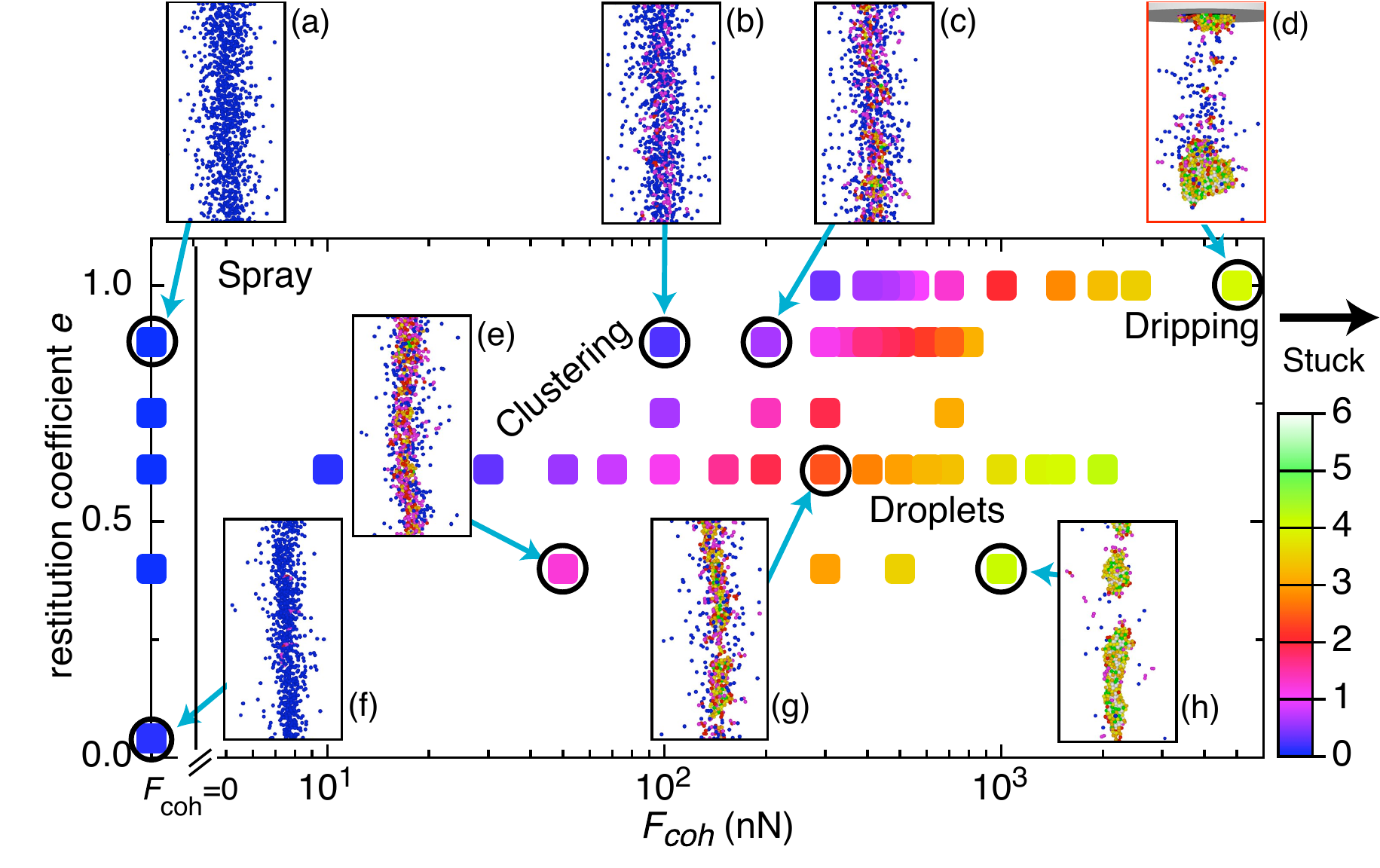}
\caption{(Color online) Phase diagram parameterized by cohesive force $F_{coh}$ and restitution coefficient $e$.  Data are for the same grain diameter and aperture size as in Fig.~\ref{fig:aspect} Data points are color coded by the average contact number $\langle C(z) \rangle$, computed at depth $z=$~4~cm.  Images (a)-(h) from the simulations at the same depth are shown for selected ($F_{coh}, e$) pairs.  Grains in these images are colored by their local contact number according to the same color scale.  Data along the left vertical axis corresponds to $F_{coh} =0$.  The image in the upper right highlighted by the red border corresponds to $z=0$ instead of 4~cm. } 
\label{fig:phase}
\end{figure*}

\begin{figure*}
\includegraphics[width=.7\textwidth]{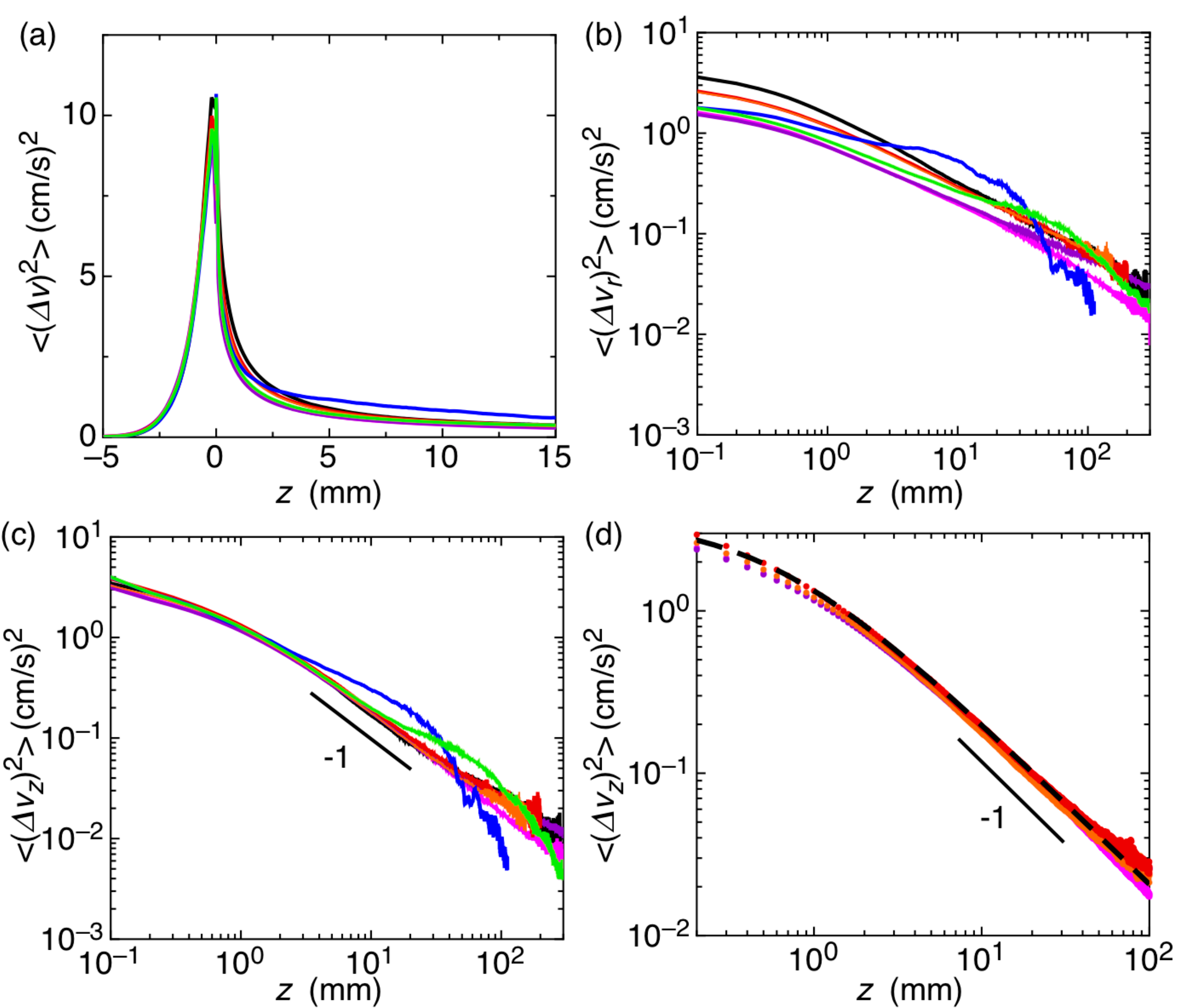}
\caption{(Color online) Average local velocity deviations as a function of depth.   Data are for the same grain diameter and aperture size as in  Fig.~\ref{fig:aspect}.  The average axial velocity at the aperture was  $v_{z,0} \approx 10$ cm/s. (a) Mean square velocity deviations $\langle (\Delta \mathbf{v})^2 \rangle $ inside the hopper ($z < 0$) and in the falling stream ($z > 0$).  (b) Radial velocity deviations 
$\langle (\Delta v_r)^2 \rangle = \langle (\Delta v_x)^2 +(\Delta v_y)^2 \rangle$.  (c) Axial velocity deviations $ \langle (\Delta v_z)^2 \rangle$.   Data in (a)-(c) correspond to: $F_{coh}=$ 0, $e=$ 0.88 (orange), $F_{coh}=$ 0, $e=$~0.48 (pink), $F_{coh}=$~50~nN,$e=$~0.40 (purple), $F_{coh}=$~100~nN, $e=$~0.88 (red), $F_{coh}=$~400~nN, $e=$~1.0 (black), $F_{coh}=$~300~nN, $e=$~0.61 (green), $F_{coh}=$~1000~nN, $e=$~0.4 (blue).    (d)  Same as in (c) with droplet forming streams (green and blue curves) excluded.  Dashed line: fit to $\Delta v_z^2$ (Eq. 2), giving $\Delta z=$~200~$\mu$m.}
\label{fig:temp_plot}
\end{figure*}

\begin{figure*}[t]
\includegraphics[width=.95\textwidth]{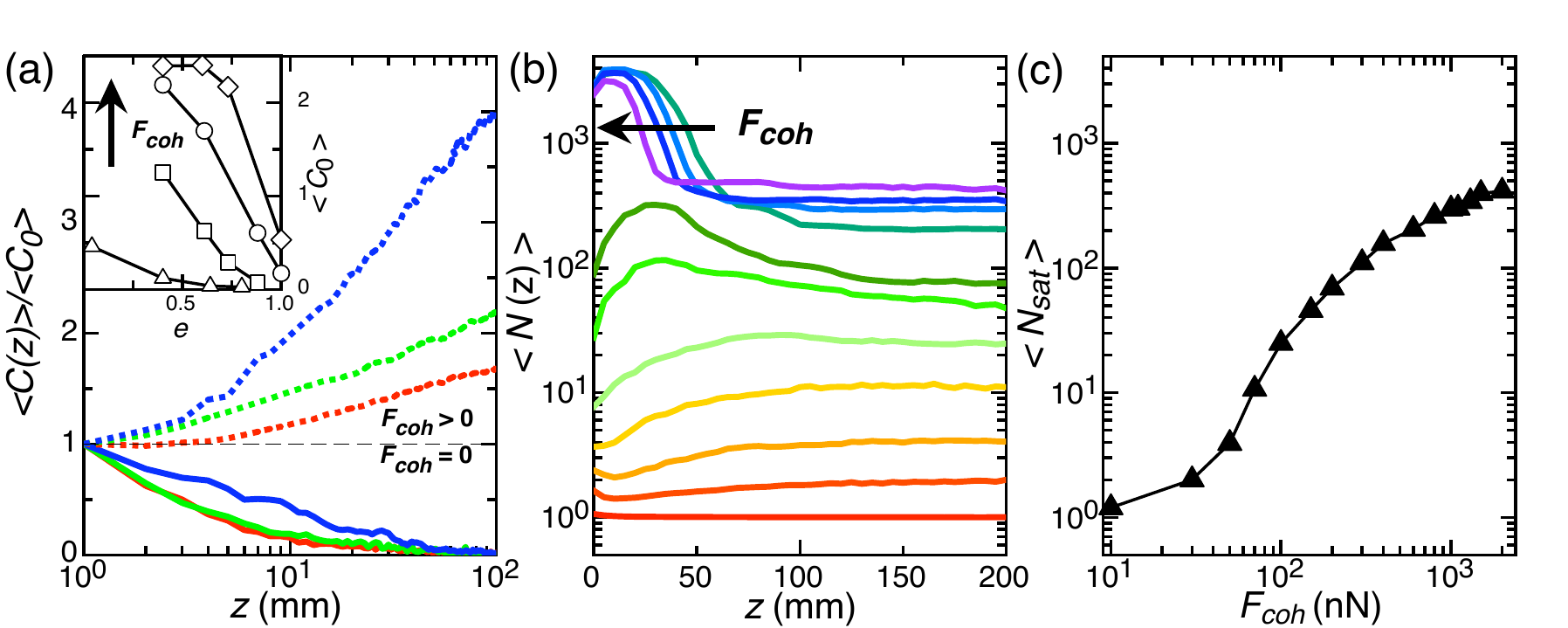}
\caption{(Color online) Evolution of contact number and network size.  Data are for the same grain diameter and aperture size as in  Fig.~\ref{fig:aspect}.  (a)  Normalized average contact number vs. distance below aperture.  Data shown for $F_{coh} =$~0~nN and $e =$~0.04 (solid blue), 0.4 (solid red), 0.88 (solid green) and for $F_{coh}=$~100~nN, $e =$~0.88 (blue dots); 300~nN, 0.61 (green dots); 1000~nN, 0.40 (red dots).  Inset: Initial contact number vs. restitution coefficient for  $F_{coh} = $~0 (triangles), 100~nN (squares), 500~nN (circles), and 1000~nN (diamonds).  (b)  Average network size vs. distance below aperture for $e=$~0.61 streams.  From bottom to top along right-hand edge: $F_{coh}=$~0~nN, 30~nN, 50~nN , 70~nN, 100~nN, 150~nN, 200~nN, 600~nN, 1000~nN, 1300~nN, 2000~nN.  (c)  Large-$z$ saturation value of average network size vs. cohesive force for $e=$~0.61 streams.}
\label{fig:contact}
\end{figure*}

\begin{figure*}[t]
\includegraphics[width=.95\textwidth]{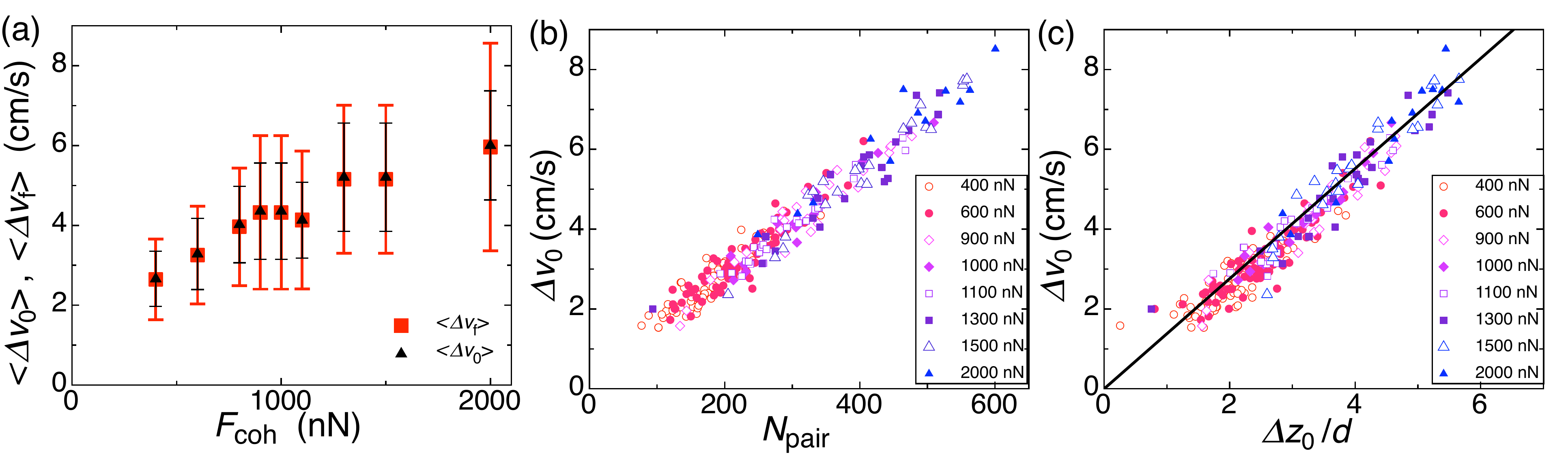}
\caption{(Color online) Initial and final separation velocity varying $F_{coh}$.  Data are for the same grain diameter and aperture size as in  Fig.~\ref{fig:aspect}. (a) Average initial separation velocity $ \Delta v_0 $ (black triangles) and final separation velocity $ \Delta v_f $ (red squares) vs $F_{coh}$. (b) $\Delta v_0 $ for pairs of adjacent droplets from streams with $e=$~0.61 and $F_{coh}$ from 400~nN to 2000~nN plotted against the mean number of grains in each pair $N_{pair} = (N_i + N_j)/2$ for droplets $i$ and $j$.  (c) $\Delta v_0$ vs droplet separation at the aperture $\Delta z_0 = \Delta z(t=0)$ scaled by the grain diameter $d$.  Solid line is a fit to $\Delta v_0 = \alpha (\Delta z_0/d)$ with $gd/v_{z,0} = $~1.4~cm/s.    } 
\label{fig:dv0}
\end{figure*}

\begin{figure}[b] 
\centering
\includegraphics[width=0.5\textwidth]{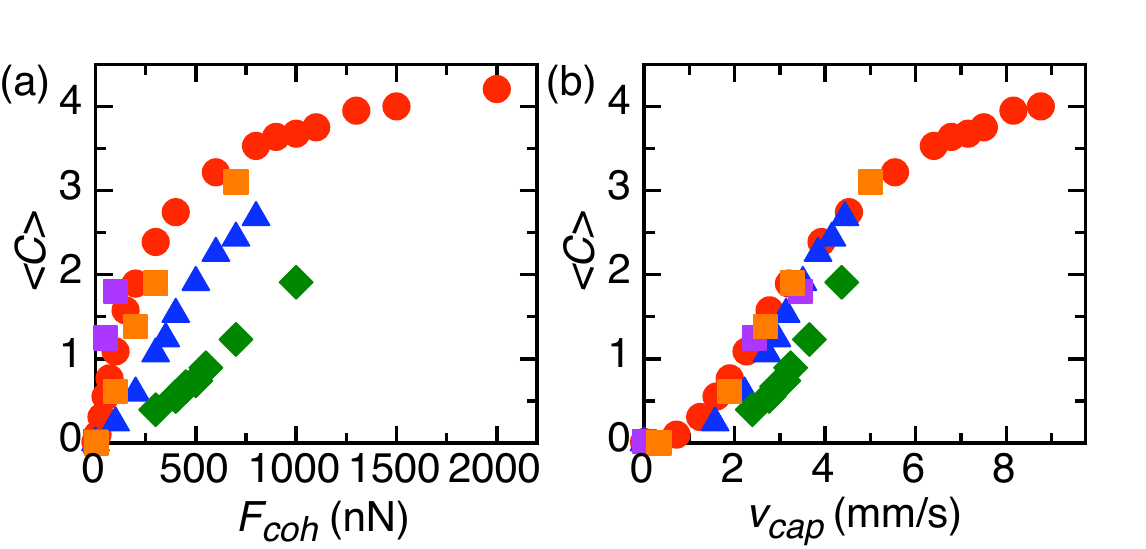} 
\caption{(Color online)  Contact number vs. $F_{coh}$ and $v_{cap}$.  Data are for the same grain diameter and aperture size as in  Fig.~\ref{fig:aspect}.  (a) Mean contact number $C$ measured $4$~cm below the aperture plotted against the cohesive force $F_{coh}$ for $e=$~1.0 (green diamonds), 0.88 (blue triangles), 0.73 (orange squares), 0.61 (red circles) and 0.40 (magenta squares).  (b)  Mean contact number at $4$~cm from the aperture plotted against the capture velocity $v_{cap}$ with same symbols as in (a).}
\label{fig:capvel}
\end{figure}

\begin{figure}
\includegraphics[width=0.44\textwidth]{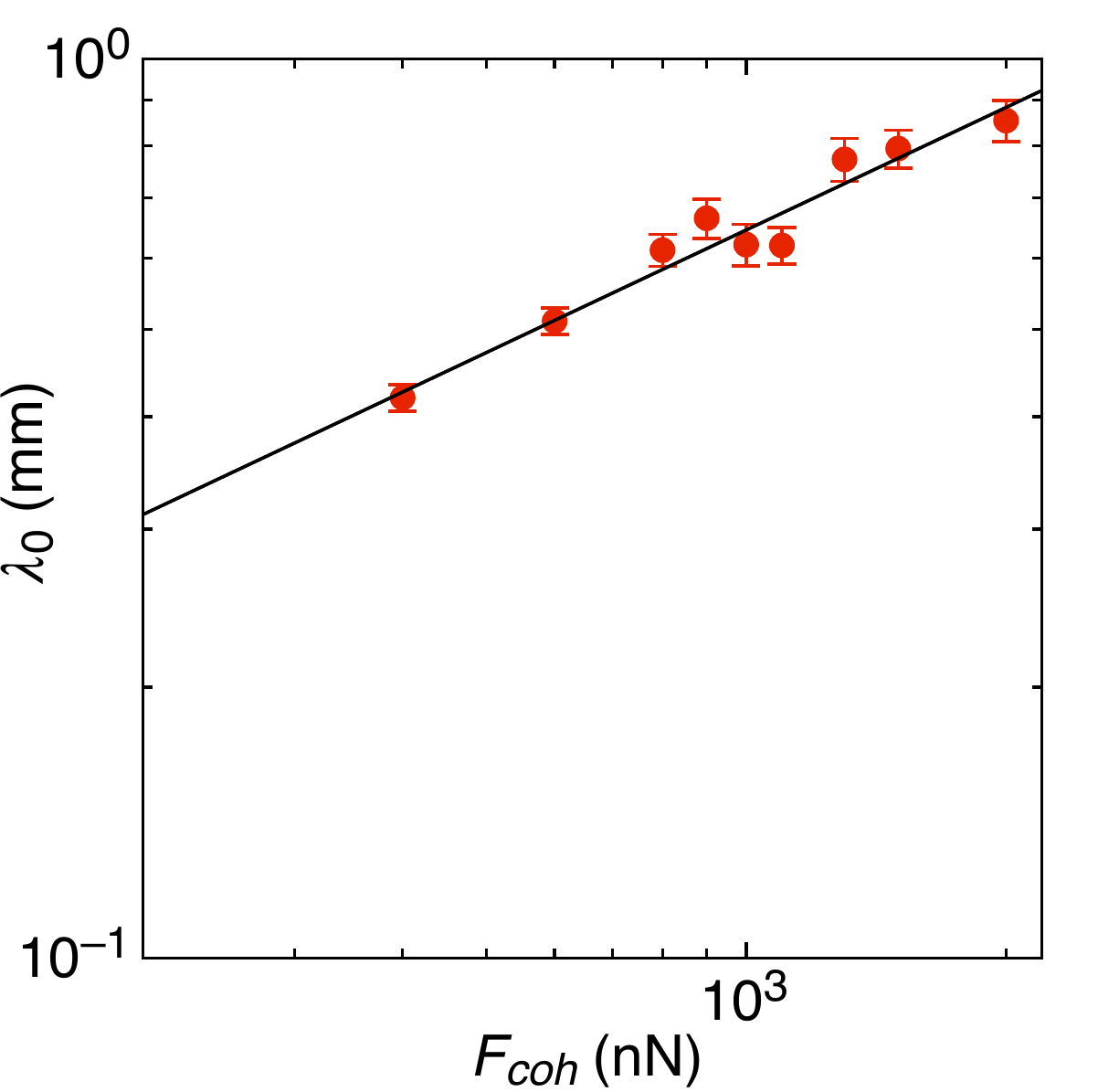} 
\caption{(Color online) Droplet lengthscale at the aperture.  Data are for the same grain diameter and aperture size as in  Fig.~\ref{fig:aspect}.  Measured droplet separation at aperture (red circles) and toy model fit (solid line).  The single fit parameter has the value $s=(2.05 \pm 0.03)\times10^{-2}$~nN/mm$^2$.}
\label{fig:model}
\end{figure}

We model the interactions between particles with molecular dynamics, including all three-dimensional degrees of freedom, following the approach described in Refs.  \cite{Brilliantov:2004p2980,CUNDALL:1979p1679,Silbert:2001p1214}. The system starts as an ensemble of up to 70,000 monodisperse spheres of diameter $d$ inside a  cylindrical reservoir (hopper).  Pulled by gravity, grains leave the hopper bottom through an aperture of diameter $D_0$, and the hopper is continually replenished at the top to keep the average fill height constant.  For practical purposes, we work primarily with $d=$~200~$\mu$m spheres and aperture size $D_0=$ 3 mm (Figs.~\ref{fig:aspect}-\ref{fig:model}), although we present data with $d=$~100~$\mu$m and $D_0=2$~mm in Fig.~\ref{fig:clusters}.  Compared to the simulations in Fig.~\ref{fig:clusters}, the smaller stream diameter to particle diameter ratio $D_0/d =$~15 for Figs.~\ref{fig:aspect}-\ref{fig:model}  is computationally more efficient, but, as in the experiments \cite{Royer:2009p41}, it leads to more scatter and less definition in the resulting stream features.  Since the larger grain mass increases the collisional kinetic energy, the strength of cohesion required for droplet formation is also increased. 

Normal and tangential contact forces are modeled with a linear spring-dashpot force law.  Collisional and frictional dissipation are parameterized by coefficients of restitution $e$ and static friction $\mu$.  Cohesion between grains is incorporated as a constant attractive force $F_{coh}$ which acts over a distance $l_c << d$.  This force only turns on after particle surfaces have come into contact.  As a result, completely separating two particles initially in contact costs an amount of energy $W_{coh}\equiv\int_{d}^{d+l_c}F_{coh}dr=F_{coh}l_c$.  This work has the same origin as $W_{ad}$ introduced by Brilliantov et al.~\cite{Brilliantov:2007p2661} in their model for viscoelastic adhesive collisions, although in our case we restrict our particles to a bare coefficient of restitution that is velocity independent and a constant cohesive force.  Physically, this model is often used to mimic short range forces such as capillary bridges and van der Waals interactions.  For a thorough discussion of these forces and their consequences, we refer the reader to the review by S. Hermingaus \cite{Herminghaus:2005p1216}.  

We do not include interactions between the particles and an interstitial fluid (air) since  experiments performed under different levels of vacuum showed that the basic features of droplet formation do not depend on the presence of air drag  \cite{Royer:2009p41}.  For the results reported here, we set $\mu = 0.5$, $l_c = 100$~nm and the specific grain density to that of glass ($\rho = $~2.5~kg/m$^3$), and we explore the effects of collisions and cohesion by varying $e$ and $F_{coh}$.

The results shown in Fig.~\ref{fig:clusters} for $d=$~100~$\mu$m grains falling from a $D_0=$~2~mm aperture reproduce the two types of behavior observed in experiments for similarly sized glass spheres with short-ranged cohesive interactions of comparable strength (as was confirmed by AFM \cite{Royer:2009p41}), namely spraying and droplet formation.  For a highly cohesive, highly dissipative stream, as shown in Fig.~\ref{fig:clusters}(a)-(d), the stream begins by narrowing in width as is commonly observed in liquid streams.  Further downstream undulations form and eventually evolve into well-defined droplets connected by necks.  These necks thin and eventually break.  When the cohesive strength was reduced in the experiments by roughening the particle surfaces (also confirmed by AFM), droplet formation ceased and the stream behaved more like a collimated spray.  The same qualitative change is reproduced when the cohesive strength is decreased in the simulations [ Fig.~\ref{fig:clusters}(e)-(h)].  Fig.~\ref{fig:clusters}(a)-(d) also shows that droplet formation goes hand in hand with the development of spatial heterogeneities in the relative particle velocities:  regions of larger relative velocity deform into necks and stay `hot' even for a short time after they break, while the interior of the droplets is `colder' and they appear to be essentially frozen once the separation process is complete [for details of this evolution see supplemental movies \cite{epaps_note2}]. 

The overall shape of the droplets can be characterized by their length to width aspect ratio $\lambda_d/w_d$.  For the purpose of establishing  shapes, height and width data was gathered for all droplets of 30 particles or more in simulations whose cohesive force $F_{coh}$ was greater than $100$~nN (see Appendix for details).  As shown in Figs.~\ref{fig:aspect}(a) and (c), most droplets have $\lambda_d/w_d$ between about one and three, in agreement with experiments \cite{Royer:2009p41}.  One feature the simulations produced that was not seen in experiments is the tail of droplets whose aspect ratios are larger than three [Fig.~\ref{fig:aspect}(c)].  These data are most likely the result of the inclusion of neighboring droplets that did not fully separate before the simulation was terminated (an observation supported by the slight increase at $\lambda_d/w_d\approx4$).  The histogram in Fig.~\ref{fig:aspect}(c) shows that $\sim75\%$ of droplets have an aspect ratio between one and three and that the most probable aspect ratio lies very near to two.  As can be seen in Fig.~\ref{fig:aspect}(b), for a given inelasticity the length and width of droplets increases with the amount of cohesion.  

These observations are in contrast with inviscid liquid streams undergoing Rayleigh-Plateau breakup.  First, liquid breakup has unstable wavelengths only for $\lambda_l/D_0 \ge \pi$ (here we use $\lambda_l$ to distinguish the liquid lengthscale from the granular lengthscale) \cite{Eggers:2008p71}.  It is important to note that the definitions of these aspect ratios are slightly different, with $D_0$ (the local unperturbed jet diameter) as the denominator in the liquid case and $w_d$ (the droplet width) in the granular case.  However, if we make the more direct comparison and calculate $\lambda_d/D_0$ for the granular case, we find that $\lambda_d/D_0\sim2/3$, drawing an even clearer distinction between the granular and liquid streams.  This comparison is still complicated by the fact that the granular stream is continually being stretched axially by gravity, and if we account for this by noting that the length of fully formed droplets is necessarily longer than the corresponding lengthscale at the aperture (which we will call $\lambda_0$), we deviate still further from the liquid case arriving at $\lambda_0/D_0\le2/3$.  Second, for inviscid liquids undergoing Rayleigh-Plateau breakup, the size of droplets is independent of the surface tension.  In fact, the only parameter governing the size of droplets is the aperture diameter, with $\lambda_l \ge \pi D_0$ determining the minimum unstable wavelength and $\lambda_l \approx 4.5 D_0$ yielding the fastest growing wavelength \cite{Eggers:2008p71}.  As seen in Fig.~\ref{fig:aspect}(b), the droplet size does depend on the cohesive strength in the granular case.  On a microscopic level, surface tension arises from cohesive bonds between constituents, whether they are molecules or granular particles, i.e. $\gamma \approx W_{coh}/d^2$ (where $\gamma$ is the surface tension, $W_{coh}=F_{coh}l_c$ is the cohesive energy stored in a bond, and $d$ is the particle diameter).  Even so, these observations make it clear that the mechanism by which this ``surface tension'' leads to systematic breakup in the granular case differs from the liquid case.

Along with cluster aspect ratio, our simulations also reproduce the observed linear growth of the separation between adjacent droplets with time [Fig.~\ref{fig:dv_dz}].  Identifying the grains which make up a fully formed droplet, we measure the vertical distance between the centers of mass for adjacent droplets $\Delta z(t)$ and compute the separation velocity $\Delta v(t) = d(\Delta z)/dt$ both in the free falling stream ($t>0$) and inside the hopper ($t<0$).  Even inside the hopper $\Delta z >0$ and $\Delta v > 0$, indicating that despite the large amount of shear there is relatively little vertical mixing.  As pairs of droplets approach the aperture they rapidly accelerate apart until they leave the aperture with a separation velocity $\Delta v_0$.  After leaving the aperture the increase in $\Delta v$ abruptly stops, though the droplets can still interact with each other, either accelerating or decelerating the separation of neighboring pairs.  Contrasting Fig.~\ref{fig:dv_dz}~(a) and (b), it becomes apparent that droplets in streams with larger cohesive forces undergo stronger, more sustained interactions, whereas droplets in the streams with the smaller cohesive force exhibit weaker, more erratic interactions.  These interactions could not be resolved in \cite{Royer:2009p41} since experimentally $\Delta z$ could only be measured at later times after well defined undulations had formed.  After a droplet and its neighbors are no longer connected to other droplets $\Delta v$ remains fixed at a final separation velocity $\Delta v_f$.

The simulations allow for the systematic investigation of parameters which are not easily changed in experiments, such as $F_{coh}$ and $e$.  Fig.~\ref{fig:phase} represents a `phase diagram' based upon these two parameters that delineates different regimes according to general, macroscopic types of stream behavior: spraying, clustering, droplet forming, and dripping.  The images provided for several of the ($F_{coh}, e$) pairs show snapshots at a depth $z= $~4.0~cm downstream.  This depth was chosen to allow sufficient time for some of the salient stream features to emerge (c.f. Fig.~\ref{fig:clusters}).  The images highlight the different regimes of stream behavior, but also reveal that it is possible for streams in different regions of phase space to appear nearly indistinguishable by shape alone, such as Figs.~\ref{fig:phase}(a)-(c) and Figs.~\ref{fig:phase}(e)-(f).  In order to tease out subtle differences in the structure and connectivity of these streams we compute each grain's contact number $C$, defined as the number of neighbors whose centers are less than $d+l_c$ away from the grain's center.  This includes both surface-to-surface contacts  and neighbors within range of the cohesive force $F_{coh}$.  For example, Figs.~\ref{fig:phase}(a) ($F_{coh} = 0$, $e=0.88$) and ~\ref{fig:phase}(b) ($F_{coh} = 100$~nN, $e=0.88$) both appear to be diffuse sprays based on the 2-dimensional projection of particle positions alone, but the color coded images reveal a substantial number of contacting pairs for the $100$ nN stream and virtually none for the zero cohesion stream.  We refer to streams that evolve into isolated particles as `sprays' and distinguish this behavior from the formation of small aggregates in the `clustering' regime.  As an example, even though a stream corresponding to $e=0.04$ and $F_{coh}=0$ [ Fig.~\ref{fig:phase}(f)] appears much denser and more collimated, there are only a few scattered contacting pairs.  In  Fig.~\ref{fig:phase}(e) ($F_{coh} = 50$ nN, $e=0.40$), however, most grains have at least one contacting neighbor and some have as many as four or five.  This already indicates that, while the inelasticity strongly affects the degree to which a stream remains collimated, a non-zero cohesive force is essential for maintaining contacts between adjacent grains.  Coloring the data points in the phase diagram by the average contact number $\langle C(z) \rangle$ at $z =$ 4 cm reveals no sharp boundaries between different regimes but rather gradual crossovers.

Towards larger cohesive forces and smaller coefficients of restitution lies the droplet forming regime [c.f.  Fig.~\ref{fig:phase}(h)] discussed already in connection with Figs.~~\ref{fig:clusters}(a)-(d). The behavior here is reminiscent of liquid flows from a small opening and individual droplets can contain hundreds of grains. If we increase the cohesion even more and approach the right-hand edge of  Fig.~\ref{fig:phase}, we enter a regime in which there is no longer a steady outflow from the aperture. In this `dripping' regime large aggregates of grains slowly ooze through the aperture [ Fig.~\ref{fig:phase}(d)].  Once the weight of these aggregates becomes too large, big chunks break off and pull away.  Further increasing the cohesive force, the grains become permanently stuck inside the hopper and all flow stops.

Since droplets and clusters are aggregates of grains with vanishing relative velocity, it is tempting to use the notion of a granular temperature to characterize the roles of inelasticity and cohesion.  In the kinetic theory of granular flows, the granular temperature is obtained by computing local deviations of grain velocities from a coarse-grained mean velocity field $\mathbf{V}(\mathbf{x},t)$ \cite{Goldhirsch:2008p1344,Goldhirsch:2003p1343}.  In freely falling streams, however, the velocity field varies strongly in both time and space.  Nevertheless, to capture the variations in relative velocities we can calculate a mean squared velocity deviation $ \langle (\Delta \mathbf{v})^2 \rangle  =  \langle (\Delta v_x)^2 \rangle + \langle (\Delta v_y)^2 \rangle + \langle (\Delta v_z)^2 \rangle $ of each grain relative to its neighbors within a small region.   Below we will loosely refer to this measure as a granular temperature, but it is important to remember that it  contains contributions from collisional velocity fluctuations and as well as from gradients in the mean velocity across a grain's neighborhood.   

In  Fig.~\ref{fig:temp_plot} we plot the mean squared velocity deviations versus depth for a sampling of parameter values from the representative regions of the phase space.  Inside the hopper and sufficiently far above the aperture, velocity deviations are small and grains move in unison.  As the aperture is approached from above, shear  `heats' up the granular medium, leading to a pronounced peak in $ \langle (\Delta \mathbf{v})^2 \rangle$.  At this point, the rms velocity deviations reach a value that is a significant fraction of the mean outflow velocity, $ \langle v_z \rangle \approx$~10~cm/s. We find that the maximum deviations are nearly the same for a wide range of cohesive forces and inelasticities, suggesting that the dynamics inside the hopper and the initial flow conditions right at the aperture are largely unaffected by changing these parameters.  

Once the particles leave the hopper the `cooling' phase begins and differences emerge.   Figs.~\ref{fig:temp_plot}(b) and (c) show separately the transverse and axial contributions to  $\langle (\Delta \mathbf{v})^2 \rangle$.  For weakly cohesive streams, and irrespective of $e$, the large-$z$ behavior is seen to approach a power law for both contributions, albeit with the exponent of the transverse component being slightly smaller.  Streams that undergo strong droplet formation (green and blue traces) show deviations from a simple power law, exhibiting first an excess of fluctuations (c.f. the shear visible in the neck regions in Fig.~\ref{fig:clusters}) followed by a rapid drop of the local droplet temperature after break-up, as cohesive forces bind grains together. 

In the absence of drop formation the observed axial cooling follows, to very good approximation, the behavior expected just from the stretching of streams of non-interacting particles, i.e., simply from the velocity gradient  due to gravity. As particles fall out of the aperture, they acquire a free fall velocity
\begin{equation}
\label{eq:v_of_z}
v_z(z) = v_{z,0}\sqrt{1+2gz/v_{z,0}^2},
\end{equation}
where $v_{z,0}$ is the axial velocity at the aperture and $g$ is the gravitational acceleration.  The axial velocity deviations over a local range $\Delta z$ are then computed as
\begin{equation}
\label{eq:dv_of_z}
\langle (\Delta v_z(z))^2 \rangle= \langle( v(z+\Delta z) - v(z)^2 \rangle.
\end{equation}
This predicts not only the $1/z$ decay of the axial temperature at large $z$, but, as seen in  Fig.~\ref{fig:temp_plot}(d) (dashed line), fits the data of non-droplet-forming streams quantitatively from the aperture on down (in these fits the initial velocity $v_{z,0}$ is known and the only free parameter is  the neighborhood size $\Delta z$).   Since $z$ is proportional to the square of the falling time, the axial temperature in  Fig.~\ref{fig:temp_plot}(d) drops as $1/t^2$, coincidentally the same behavior as predicted for the early stages of a non-expanding granular gas \cite{Ulrich:2009p2669}.  We note that this differs from  the $1/z^2$ decay proposed by Amarouchene {\it et al.} \cite{Amarouchene:2008p2673} based on fluctuation measurements at the surface of streams falling in air.

While the mean squared velocity deviations reveal the important role played by gravity in stretching and thereby diluting and cooling the stream,  they are not particularly sensitive to details introduced by inelasticity or cohesion. We see this directly in  Fig.~\ref{fig:temp_plot} where temperature curves for a range of $F_{coh}$ and $e$ values effectively collapse on top of each other except when the stream undergoes gross changes in the behavior, such as the onset of large droplets. It is therefore important to explore additional and possibly more useful indicators of the roles of  $F_{coh}$ and $e$.

One of these indicators is the depth dependence of the average contact number, $\langle C(z) \rangle$.  To emphasize changes relative to the initial particle configuration near the aperture, we normalize $\langle C(z) \rangle$ by the average initial contact number, $\langle C_0 \rangle$, measured at $z=1$~mm. For cohesive streams in the clustering, droplet forming, or dripping regimes, the normalized contact number always increases with $z$ [ Fig.\ref{fig:contact}(a)].  Contrarily, in non-cohesive streams, $\langle C(z) \rangle/\langle C_0 \rangle$ always decreases, independent of the inelasticity and even in streams with restitution coefficients approaching zero.  Since the interparticle distance, and hence the collision rate, in the stream steadily decreases with depth, the usual pressure gradients associated with inelastic collisions are rendered less effective, and there is no restoring force to keep contacting grains together. As a result, even the slightest separation velocity--essentially ensured by the velocity gradient--will eventually cause a cluster of contacting grains to drift apart.   Only in cohesive streams is it possible to form particle aggregates that will remain stable against small velocity differences in the absence of further collisions.  

While the normalized average contact number draws a clear distinction between cohesive and non-cohesive streams, it does not differentiate cleanly between the more subtle consequences of cohesion.  Such differences are highlighted when we go beyond nearest neighbor contacts and look at the full network of particles connected to a given grain.  We do this by defining the network size $\langle N(z) \rangle$ as the average number of grains connected to a grain at a distance $z$ below the aperture (see Appendix for averaging procedure).  We define two grains as connected if a path exists between them that consists of pairs of grains whose center-to-center distances are no larger than $l_c+d$.   Fig.~\ref{fig:contact}(b) shows the results for streams with $e=0.61$ and a range of $F_{coh}$ values. 

The behavior of $\langle N(z) \rangle$ can be classified into three basic categories.  For $F_{coh}=0$ we find that $\langle N(z) \rangle$ monotonically decreases toward unity, in agreement with the conclusions drawn from the evolution of $\langle C(z) \rangle$. This is the spraying regime.  For $F_{coh}$ less than about $\sim$~100~nN, $\langle N(z) \rangle$ first dips sharply as particles leave the hopper, but then undergoes a prolonged period of slow growth, after which it saturates at values less than $\sim$~30 grains.  We interpret this as the signature of clusters that form by collide-and-capture events.  The dynamics in this regime consist of an intricate interplay between the effects of the velocity profile at the aperture, the stretching and diluting of the stream via gravity, and low energy collisions which permit initially separated grains to become permanently connected.  The capture process can start once the stream has fallen and cooled to a low enough level such that the cohesive energy $F_{coh}l_c$ exceeds the pre-collisional kinetic energy \cite{Spahn:2004p2498, Ulrich:2009p2669}.  This does not continue indefinitely, however, because the stretching due to gravity drives the collision frequency toward zero.  This situation is remarkably similar to the trade-off between cooling and dilution observed in free expansion experiments on cryogenic He jets \cite{Bruch:2002p2686}.  

For larger $F_{coh}$, between 100~nN and 2000~nN, the behavior reverses. Now $\langle N(z) \rangle$ quickly increases to a peak value before settling down to its asymptotic value at large $z$.  This is the droplet regime. The initial increase in  $\langle N(z) \rangle$ is due to the liquid-like narrowing of the stream below the aperture, where the associated radial influx and sufficiently large cohesive energy work together to favor network growth.  The subsequent decrease signals the break-up of the stream into droplets.  The initial network growth for the strongly cohesive streams becomes smaller as the cohesive force is increased.  At the same time, the peak network size saturates at some maximum value ($\sim$~1100 grains).  We also see that the break-up is more pronounced in more cohesive streams, i.e. the sudden drop in $\langle N(z) \rangle$ occurs more rapidly [inset to  Fig.~\ref{fig:contact}(b)].  Further inspection of  Fig.~\ref{fig:contact}(b) reveals that the location of this dropoff moves up toward the aperture as  $F_{coh}$ is increased, suggesting a natural progression from droplet forming to dripping streams.  

The qualitatively different shapes of the $\langle N(z) \rangle$ curves reveal a fundamental difference between clustering and droplet forming streams.  A clustering stream starts as `gas' of individual grains or collections of small network fragments which grow larger by capturing additional particles during collisions. In contrast, a droplet forming stream starts as large network of grains that eventually breaks into smaller portions, again similar to the formation of drops in liquid He jets \cite{Harms:1997p2674}.  A potential crossover between these two modes can be identified by tracing the asymptotic, large-$z$ limit of $\langle N(z) \rangle$ as a function of cohesive strength.  In  Fig.~\ref{fig:contact}(c) we plot the saturation network size $\langle N_{sat}\rangle$, defined as $\langle N(z=$~20~cm~$) \rangle$ vs.  $F_{coh}$.  Clustering streams undergo accelerated growth in $\langle N_{sat} \rangle$ as $F_{coh}$ is increased, (e.g. before about $\sim 100$~nN) while droplet forming streams exhibit a slow turnover in which the growth rate decreases.  If we inspect Fig.~\ref{fig:contact}(b) more closely, we see that the curves undergo the qualitative change from `slow growth' (clustering) to `eventual breakup' (droplet formation) around $\sim 100$~nN as well.   

In Fig.~\ref{fig:dv0}(a) we examine how $F_{coh}$ affects the dynamics of the stream breakup in the droplet forming regime.  The average separation velocity between adjacent droplets exhibits a weak dependance on $F_{coh}$.  The averages of the initial  $\langle \Delta v_0 \rangle$ and final $\langle \Delta v_f \rangle$ separation velocities are essentially identical, though the distribution of $\Delta v_f$ is considerably wider due to the interactions between droplets as they separate.   When $\Delta v_0$ is plotted against the mean number of grains in individual pairs, $N_{pair}$,  the data for different $F_{coh}$ collapse onto a single curve [Fig.~\ref{fig:dv0}(b)], indicating that the dependence of $\langle \Delta v_0 \rangle$ on $F_{coh}$ stems entirely from the increased network size at larger cohesive forces.  The data similarly collapse when plotted against the center of mass distance between the clusters at the aperture $\Delta z_0 = \Delta z(t=0)$  [Fig.~\ref{fig:dv0}(c)].  From Eq.~\ref{eq:v_of_z}, the velocity difference of two points in the stream near the aperture and separated by a distance $\Delta z_0 \ll v_{z,0}^2/g$ is
\begin{equation}
\label{eq:dv0}
\Delta v_0  \simeq \frac{gd}{v_{z,0}}\bigg{(}\frac{\Delta z_0}{d}\bigg{)}.
\end{equation}
The fit shown in Fig.~\ref{fig:dv0} yields $gd/v_{z,0} = $~1.4~cm/s, reasonably close to the value 1.7 cm/s predicted from just the free-fall velocity profile.  This shows that the droplet separation is primarily driven by the free-fall velocity gradient along the stream and is set by the initial droplet separation, which is on the order of a few grain diameters.  This scale is in agreement with the extrapolated value of the length scale measured in previous work on granular streams \cite{Mobius:2006p566}.   

So far, the simulations have allowed us to reproduce experimental results and access data that would otherwise be impossible to measure.  We can also use them to test possible mechanisms for the regions in phase space outlined in Fig.~\ref{fig:phase}.  For the capture phase, it is natural to simply think of the stream as a gas of colliding grains.  In this scenario, structure formation takes place mainly through collide-and-capture events.  The picture is simplified if we assume all collisions are binary.  As pointed in \cite{Brilliantov:2007p2661}, the minimum relative pre-collision velocity $v_{cap}$ two particles must have in order to separate after a collision is
\begin{equation}
v_{cap}=\sqrt{\frac{2\,l_c\,F_{coh}}{m_{\text{eff}}e^2}}.
\end{equation}
If details regarding the density and the distribution of relative particle velocities at the aperture are fairly constant, it is natural to expect that the mean contact number $C$ will depend primarily on $v_{cap}$, i.e. $F_{coh}$, $l_c$, and $e$.  This is tested in Fig.~\ref{fig:capvel}.
Indeed, plotting $\langle C \rangle$ vs. the capture velocity as opposed to the cohesive force does bring the data together appreciably even for higher force streams, but it does not quite capture the whole story.  Streams with lower $e$ systematically deviate to higher mean coordination numbers.  This observation is consistent with the fact that the introduction of a cohesive force leads to an effective coefficient of restitution \cite{Brilliantov:2007p2661}, in our case given by
\begin{equation}
e_{eff}=e\sqrt{1-\left(\frac{v_{cap}}{v_{rel}}\right)^2}.
\end{equation}
It follows that even if two dense granular flows consist of particles with the same capture velocity, the one with the higher $e$ (and consequently $e_{eff}$) should exhibit more structure formation.  

Deep in the droplet forming regime, it is tempting to look for a lengthscale that falls out from our free parameters since the size of droplets scales with the cohesive energy.  We can begin to picture a mechanism for the selection of this lengthscale by recalling two observations.  The first is that well into the droplet regime virtually all grains leaving the aperture are connected in one large network.  The second is that the velocity gradient at the aperture plays an important role.  Putting these together, one might suspect that droplet formation is closely related to the kinetic energy imprinted on the stream by the gradient.  As a simplified model, we assume that our stream is a well-connected column of grains of infinite length and diameter $D_0$, isolated from all external forces and subject to an initial axial velocity gradient of constant value $\dot{\gamma}_0$.  We partition this stream into segments of length $\lambda_0$ which will eventually become droplets.  For an individual droplet, conservation of energy requires
\begin{equation}
	K_i+W-U_i=K_f+\Delta E_{int},
\label{eq:delta_K}
\end{equation} 
where $K_i$ is the droplet's initial kinetic energy, $W$ is the work done on it by neighboring droplets, $U_i$ is the initial energy stored in bonds ($U_i \ge 0$), $K_f$ is the final kinetic energy, and $\Delta E_{int}$ is the change in the internal energy of the droplet.  We can calculate the initial kinetic energy in the rest frame of the droplet from our knowledge of the initial velocity gradient
\begin{equation}
	K_i=\int_{\lambda_0/2}^{-\lambda_0/2}\frac{1}{2}\rho\phi_0\pi\bigg{(}\frac{D_0}{2}\bigg{)}^2(\dot{\gamma}_0 z)^2dz\equiv \epsilon \lambda_0^3,
\label{eq:K_i}
\end{equation}
where $\rho$ is the specific density of the grains and $\phi_0$ is the initial packing fraction. In this model with identical droplets, the forces on a droplet from the right and left must be equal throughout the entire process, so its center of mass velocity must remain constant.  Though not strictly the case in the simulations where the droplet size can vary, on average this is true, as evident from Fig.~\ref{fig:dv0}(a).   We assume that $K_f$ is identically zero, i.e. no energy remains in the form of particle vibrations or bulk rotations.  

The work $W$ is more difficult to estimate because it depends strongly on the dynamics of the droplet formation process.  We know it is present and non-negligible by the separation dynamics shown in Fig.~\ref{fig:dv_dz}, which also suggest that this work grows with $F_{coh}$.  We can estimate its magnitude by noting that it is just the total force in the cross section between the droplets, $F_{cs}(t)$, integrated over the effective length which the force pulls.  In the rest frame of our droplet, this work is approximately
\begin{equation}
 W\approx2\int_{t=0}^{t=t_b}F_{cs}(t) \frac{dl}{dt} dt=\dot{\gamma}_0\lambda_0\int_{t=0}^{t=t_b}F_{cs}(t)dt \equiv \alpha \lambda_0,
 \label{eq:Work}
\end{equation}
where $\alpha$ depends on the time to breakup $t_b$, the velocity gradient $\dot{\gamma}_0$, and the cohesive force $F_{coh}$.  The assumption of a constant velocity gradient is not perfect, since interactions with neighboring droplets and internal dissipation can, in principle, alter it.  However, we have seen from Fig.~\ref{fig:temp_plot}(d) that even droplet forming streams initially exhibit the same $1/z$ decay in the mean square velocity deviations seen in non-droplet forming streams, indicating that deviations from a constant gradient are only significant close to break up.  The integral $I=\int_{t=0}^{t=t_b}F_{cs}(t)$dt depends on the full evolution of the bond configurations between the neighboring droplets, making it difficult to evaluate.  However, from Figs.~\ref{fig:aspect}~and~\ref{fig:dv_dz} we see that the droplet width, length and break up time vary only weakly with $F_{coh}$, suggesting that these changes are minor and to lowest order $I \propto F_{coh}$.  In Fig. \ref{fig:dv0} we demonstrated that the initial velocity gradient $\dot{\gamma}_0$ is independent of $F_{coh}$.  Putting these observations together suggests that, to lowest order, $\alpha \propto F_{coh}$. 

The last quantities we must account for are $U_i$ and $\Delta E_{int}$.  These are closely related, and to account for them we start with a simple example.  Imagine a single droplet (with no neighbors) of length $\lambda_0$ and initial axial velocity gradient $\dot{\gamma}_0$.  The particles in this droplet are perfectly elastic and frictionless.  Assuming the total cohesive energy remains constant, the droplet can remain intact if the energy stored in bonds exceeds the initial kinetic energy, i.e. $U_i =N_bW_{coh}\ge K_i = \epsilon \lambda_0^3$, where $N_b$ is the total number of bonds in the drop.  The number of bonds $N_b$ should grow linearly with the droplet length ($N_b=n_b\lambda_0$), which implies that the maximum length our droplet can be in order to remain intact is $\lambda_0=\sqrt{n_b^0W_{coh}/\epsilon}$.  Since we assumed no bonds are broken, energy is conserved and the particles will exchange kinetic for potential energy as bonds are `stretched' and retracted.  If we now assume that during this process the number of bonds actually grows [as it does in Fig.~\ref{fig:contact}(a)], our stream is able to counteract even more kinetic energy (note that creating new bonds adds no kinetic energy since our force is hysteretic).  We also expect that the number of bonds gained should scale linearly with the length.  Thus our droplet length can now grow to $\lambda_0=\sqrt{(n_b^0+n_{bg})W_{coh}/\epsilon}$, where $n_{bg}$ is the number of bonds gained per unit length.  As more bonds are gained, the droplet finds itself in a deeper effective potential well and must have more kinetic energy--i.e. more length--to escape.  The same is true if we add dissipation in the form of friction or inelasticity.  For each dissipative term added, more kinetic energy can be extinguished and this produces longer droplets.  The work done by neighboring droplets has the opposite effect, reducing the maximum stable $\lambda_0$.  However, for droplet formation we must always have $U_i+\Delta E_{int} > W$.  The local rearrangements and deformations that lead to this dissipative term are easily seen in our supplemental movie \cite{epaps_note2}, where we show the evolution of a single droplet.  A similar phenomenon occurs in liquids, where increasing the viscosity leads to larger droplet masses in the slow dripping from a tube \cite{Wilson:1987p561}.    

Assuming the contact distribution and collision dynamics are uniform across the length of the droplet, dissipation from friction and inelasticity should also scale with the length of the droplet, so that we can write the sum of $U_i$ and $\Delta E_{int}$ as
\begin{equation}
U_i+\Delta E_{int} = \beta \lambda_0,
\label{eq:E_loss}
\end{equation}
where $\beta$ includes contributions from all initial and gained bonds, friction and inelasticity.  Putting our proposed forms for $K_i$, $W$, and $U_i+\Delta E_{int}$ into Eq.~\ref{eq:delta_K}, we arrive at an expression for the initial length of droplets:   
\begin{equation}
  \lambda_0=\left( \frac{\beta-\alpha}{\epsilon} \right)^{1/2}.
  \label{eq:pre_model}
\end{equation}

Although precisely predicting $\lambda_0$ requires analytical expressions for $\alpha$ and $\beta$ as a function of all our simulation parameters and initial conditions, we can make some progress by just looking at the scaling with $F_{coh}$.  We have already argued that $\alpha$ (and consequently $W$) and $U_i$ scale linearly with $F_{coh}$.  Since there is no structure formation for $F_{coh}=0$, we expect that $\Delta E_{int}$ (and consequently $\beta$) also scales linearly with $F_{coh}$ to first order.  This implies
\begin{equation}
  \lambda_0=\left( \frac{F_{coh}}{s} \right)^{1/2},
\label{eq:model_fit}
\end{equation}
where $s$ is a constant with units of stress.

This toy model makes a number of important simplifications.  Foremost is the fact that near the aperture the `real' streams are more complicated than we have assumed.  Much of this results from the presence of radial velocity gradients (both $dv_z/dr$ and $dv_r/dr$), which change the initial kinetic energy and the initial rate at which energy is lost. 

Nonetheless, in comparing our idealized model with data from the simulations we find good agreement.  In Fig.~\ref{fig:model} we plot the droplet lengthscale measured at the aperture as a function of the cohesive force along with a fit to Eq.~\ref{eq:model_fit} with $s$ as a free parameter.  As can be seen, the $F_{coh}^{1/2}$ scaling of the model is consistent with behavior found in the simulations.  We see that at $\sim100$ nN, the initial lengthscale is approximately equal to the particle size ($200$~$\mu$m).  Intuitively, one might expect this model to break down once the initial lengthscale drops below the particle size, and the fact that the force at which this occurs is very near to the crossover identified in Fig.~\ref{fig:contact} is highly suggestive.  The fit corresponds to $s=(2.05 \pm 0.03)\times10^{-2}$ nN/mm$^2$.  Referring back to Eqs.~\ref{eq:pre_model} ~and~\ref{eq:model_fit}, we see that this quantity encapsulates the ratio of the initial kinetic energy $K_i$ to the net energy lost $\Delta E_{int}-W$ (divided by factors of $F_{coh}$ and $\lambda_0$ to give the appropriate units).  Physically, it is a proxy for the stream's effectiveness to counteract its initial kinetic energy through cohesion and dissipation.  While it would be preferable to have a larger range of data to fit to, this is not possible in our stream geometry.  As can be seen in Fig.~\ref{fig:phase}, the droplet forming regime gives way to the dripping regime quite rapidly just beyond the maximum force used in Fig.~\ref{fig:model}, and we are left with only about a decade in $F_{coh}$ and $\lambda_0$ before a new mechanism comes into play.  
 
\section{Conclusions}
Our results show that freely falling granular streams provide a highly sensitive probe of minute, local-level interactions between granular particles.  Gravity stretches and cools the granular stream, similar to what happens during the expansion phase in molecular jets. This makes it possible to distinguish between situations with and without attractive particle interactions, irrespective of the coefficient of restitution.  Attractive interactions that are weak (nanoNewtons), short-ranged (tens to hundreds of nanometers) and thus hard to quantify {\it{in-situ}} with other techniques are easily detected by tracking the formation of particle clusters or droplets.  In particular, our findings demonstrate that the average number of particles in a cluster or droplet, given by the saturation network size, is directly related to the cohesive strength.  This also implies a simple experimental means for detecting small amounts of interparticle cohesion without the need to follow the stream in the co-moving frame:   From images taken in the large- z limit the number of particles per droplet can be estimated, providing an upper bound on the cohesive energy and making these streams a potentially useful tool for other areas of granular physics where cohesive forces are known to play important roles but are difficult to quantify, such as industrial fluidized beds \cite{Weber:2006p2638} and accretion in protoplanetary disks \cite{Blum:2008p21}.

Nearly all aspects of stream behavior are affected by the free-fall velocity profile imparted on the stream at the aperture.  This observation suggests that the breakup mechanism itself may be a consequence of this gradient.  With this in mind, we developed a toy model in which droplets are formed as the kinetic energy associated with the gradient and the work done by neighboring droplets competes with the energy stored and bonds and lost to dissipation.  This model predicts that adjacent droplets should separate with constant velocity, which is confirmed in the simulations.  Using this model to determine the length of droplets requires some knowledge of how the energy is transferred between the initial and final states.  Predicting this is difficult because energy is lost to inelasticity, friction, and the rearrangement of bonds, all of which depend on the microscopic dynamics during the breakup process.  Nonetheless, we expect that, to first order, these terms should just scale linearly with the cohesive force.  This results in the initial droplet lengthscale scaling like $F_{coh}^{1/2}$.  This prediction is consistent with our data for droplet forming streams in the range of data available to our stream geometry.    

Our results indicate that in two very important ways granular breakup differs from inviscid Rayleigh-Plateau breakup.  First is the fact that in the liquid case the lengthscale over which breakup can occur is independent of the strength of the interaction; surface tension is only required to start the process.  For granular breakup, our results show that the size of droplets grows with the cohesive energy, indicating a competing mechanism is present.  Second, in granular breakup dissipation plays a primary role, while in liquid breakup dissipation is not even necessary.  In inviscid liquid breakup, droplets actually end up with more kinetic energy than they start with.  This is a consequence of energy conservation; in order for a liquid stream to reduce its surface energy, its kinetic energy must increase.  In granular breakup, it is clear that fully formed droplets have virtually no kinetic energy in the co-moving frame.

In a liquid stream, it is the thermal nature of the molecules which allows them to explore phase space and ultimately reduce the surface energy of the stream.  Macroscopic granular systems, on the other hand, are athermal, and if we imagine a well-connected column of grains with no macroscopic velocity gradients, we would never see breakup.  In the presence of a velocity gradient, grains have the opportunity to rearrange in a systematic way that maintains high density, allows access to energetically favorable configurations (by increasing the number of bonds), and leads to energy loss via dissipation.  In this sense, the kinetic energy from the velocity gradient in a granular stream plays a similar role to the thermal energy in a liquid stream in that it provides a way for the stream to explore otherwise unaccessible configurations.  However, in the absence of external energy input this kinetic energy is quickly lost as the grains explore new configurations and is essentially zero once the drops separate.  Combined with the extremely low effective surface tension, this prevents the granular droplets from  assuming the smooth spherical shape of liquid drops. 

While our model and predictions can account for many of the trends and behaviors seen in both simulations and experiments, they also raise a number of important questions.  One of these is how the break-up is changed, or if it even occurs, with other interactions such as conservative potentials or frictionless particles, since changing these will significantly affect how the particles can redistribute their kinetic energy.  Also, while our model produces scaling for $F_{coh}$ consistent with the simulations, it is not able to determine what happens between the initial and final states, nor is it able to explain why the aspect ratios of fully formed droplets are largely independent of the stream parameters.  As a final question, one has to wonder how the initial, local particle packing inside the stream near the aperture affects the breakup since the collision dynamics and the stream's ability to create new bonds will be strongly affected by this parameter.  

Finally, the fact that already minute amounts of cohesion can give rise to cluster or droplet formation provides a new perspective for interpreting other recent experiments.  This includes the liquid-like breakup of highly collimated jets that rise up vertically after a heavy object impacts fine, loose granular material \cite{Lohse2005p3045} and the clustering inside thin granular sheets that form after a dense granular stream hits an obstacle \cite{Cheng:2007p2753}.   As in the granular streams, heterogeneities become visible while the jets or sheets expand after a brief stage of intense collisional interactions.  Inelasticity was suspected as the source, but our findings show instead that the observed dynamic structure formation is indeed the vestige of residual attractive interactions. 

\section{Appendix}
Here we describe the technical aspects of the simulations in more detail.   Three-dimensional molecular dynamics simulations of both hopper and stream are performed using Itasca PFC3D software.  The system consists of a cylindrical hopper (diameter $D_H$) with a circular aperture (diameter $D_0$) at the center of its base.  The hopper geometry is designed to produce constant flow with the smallest number of grains necessary \cite{Nedderman:1982p1597}.  Maintaining $D_H\gtrsim2.5D_0$, $D_H\gtrsim30d+D_0$, and a fill height $H\gtrsim1.2D_H$ insures particle flow that is insensitive to the addition of further grains.  To establish the proper fill height,  the simulations maintain $N_{hop}$ particles inside the hopper.  Values for the simulation parameters used are given in the table below. For $d, D_0, D_H$ and $N_{hop}$ the first value listed refers to the streams shown in Fig.~\ref{fig:clusters} and the second to the images and data shown in Figs.~\ref{fig:aspect}-\ref{fig:model}. To initiate a simulation, grains are rained into the hopper and allowed to settle under gravity into a granular bed of height $\approx H$ before the aperture is opened (incidentally, we believe that the systematic deviations in the $e=0.61$, $F_{coh}=1000$~nN,~$1100$~nN streams may have their source in insufficient system initialization).  Once the flow starts, grains are added to the top of the bed to keep $N_{hop}$ and therefore $H$ constant on average.  The average particle velocities at the aperture produced by the simulations closely matched those found in experiments \cite{Royer:2009p41} with similar geometries (e.g., $v_z \approx $~10cm/s  for $d = $200~$\mu$m and $D_0 =$3~mm).

\begin{table}
\caption{Simulation parameters}
\label{table:parameters}
\begin{tabular}{ll}
Parameter &   Value \\
\colrule
$d$ 			& $100\,\mu\mathrm{m}$, $200\,\mu\mathrm{m}$  	\\
$\rho$		& $2.5\, \mathrm{kg\, m}^{-3}$ 					\\
$D_0$ 			& $2.0\,\mathrm{mm}$, $3.0\,\mathrm{mm}$		\\
$D_H$			& $5.0\,\mathrm{mm}$, $7.5\,\mathrm{mm}$      \\
$N_{\text{hop}}$ 	& $7$x$10^4$, $3$x$10^4$				\\
$k_n$, $k_t$ 		& $500\,\mathrm{N}/ \mathrm{m}$			\\
$\mu$ 			& $0.50$							\\
$l_c$ 			& $100\,\mathrm{nm}$					\\
\end{tabular}
\end{table}

The repulsive force between grains is modeled using a linear spring-dashpot with both normal and tangential damping as well as static friction, following \cite{Silbert:2001p1214}.  In this model, the normal contact force is given by $F_n = k_n \delta_n + \gamma_n m_{eff} \dot{\delta_n}$, where $\delta_n=d-|\mathbf{r}_i-\mathbf{r}_j|$ is the overlap between the grains $i$ and $j$, $k_n$ the normal stiffness, $\gamma_n$ is the damping constant for normal motion, $m_{eff} = m/2$ is the reduced mass, and $m = \frac{1}{6}\pi\rho d^3$ is the grain mass ($\rho = $2.5~kg/m$^3$).    The tangential contact force $F_t$ has the same functional dependence, but with tangential displacement $\delta_t$, tangential stiffness $k_t$ and damping constant $\gamma_t$ in place of $\delta_n$, $k_n$ and $\gamma_n$.  Friction is incorporated by truncating $F_t$ in order to satisfy the yield criterion $F_t \leq \mu F_n$.   

Head-on collisional energy loss was parameterized by a coefficient of restitution $e_n \equiv v_n'/v_n$, where $v_n$ and $v_n'$ are the normal components of the colliding grains' relative velocities immediately before and after the collision.   This linear-dashpot model leads to a  restitution coefficient $e_n =\exp{[-\pi \gamma_{n} / \sqrt{ 4k_{n}/m_{eff}-\gamma_{n}^2}]}$, independent of impact velocity (note that we define $e_n$ as a `bare' restitution coefficient that does not take into account cohesion).   Similarly, a coefficient of restitution for the tangential motion $e_t$, with $k_t$ and $\gamma_t$ replacing $k_n$ and $\gamma_n$, captures losses due to sliding.   For the simulations presented here, we set $k_n = k_t$ and  $\gamma_n = \gamma_t$ so that there is the same coefficient of restitution $e_n = e_t = e$ for normal and tangential motion.  The elasticity is varied by changing only the damping constants $\gamma_n$ and $\gamma_t$ while keeping $k_n$, $k_t$, and $m_{eff}$ constant. 

Cohesion is included by adding a force of constant magnitude $F_{coh}$ that turns on after two grains have collided and drops to zero outside a thin shell of thickness of $l_c$ around each grain.  Physically, this model represents a force that turns off over a lengthscale that is much smaller than the particle size, e.g. van der Waals interactions or the rupture of a liquid bridge \cite{Herminghaus:2005p1216}.   As a result, this is a hysteretic interaction and completely separating two grains initially touching costs a cohesive energy $W_{coh} = \int_d^{d+l_c} F_{coh} dr=F_{coh}l_c$.  We note that our model for cohesion between grains does not capture more subtle, higher order effects caused by, for example, a velocity dependent coefficient of restitution [see for example \cite{Brilliantov:2007p2661}].  

Droplet widths and heights are calculated for networks with 30 or more particles in streams with $F_{coh} \ge 100$~nN.  Effects of stream initialization (e.g. plug formation) are avoided by excluding the 1000 droplets farthest from the aperture.  To ensure droplets are fully formed/comopletely separated, only networks whose center of mass locations are greater than 20~cm from the aperture are included. The droplet height is defined as the maximum extent of a network along its principle axis.  The droplet width is taken to be the mean radial extent from the principle axis of each droplet averaged over all angles (partitioned into $36\,^{\circ}$ bins).  

In our system, the heterogeneous nature of the stream made it difficult to define a coarse grained velocity field.  We instead define the mean squared velocity deviations $\langle (\Delta \mathbf{v})^2 \rangle$ for each grain as the standard deviation of its velocity relative to the velocities all other neighbors within a small region, which we set at 1.5$d$.   If a grain has less than three neighbors within this region it is excluded, thus ensuring that isolated grains do not contribute.  The velocity deviations are calculated for each grain at time steps 0.2 ms apart, and then binned into 100~$\mu$m wide bins along the axial direction to make the plots in  Fig.~\ref{fig:temp_plot}.

To calculate the normalized average contact number and average network size, we define grains to be in contact if the separation between their surfaces was less than the thickness of the cohesive shell $l_c$.  The data in  Fig.~\ref{fig:contact}(a) are calculated for each grain for intervals $0.2$ ms apart and then binned into 1~mm wide along $z$.  Data in  Fig.~\ref{fig:contact}(b)-(c) were also calculated for each grain in intervals $0.2$~ms apart but binned into 5~mm wide bins along $z$.

\begin{acknowledgments}

We thank Sid Nagel, Tom Witten, Wendy Zhang, Eric Corwin, Marc Miskin, Ali Koser, and Suomi Ponce Heredia for insightful discussions.  We would also like to thank our anonymous referees for helpful suggestions.  This work was supported by the NSF through CBET-0933242.  We acknowledge use of shared facilities provided by the Keck Facility for Ultrafast Imaging at the University of Chicago and by the Chicago MRSEC through NSF DMR-0820054.  Software support by Itasca Consulting Group, Inc., under the Itasca Educational Partnership is gratefully acknowledged.  HG thanks the German-American Fulbright Commission for fellowship support during his stay at the University of Chicago.

\end{acknowledgments}

\bibliographystyle{apsrev}

%\bibliography{droplets}

%\end{thebibliography}

\end{document}